\documentclass[aps,superscriptaddress,preprintnumbers,showpacs]{revtex4}
\usepackage{amssymb}
\usepackage{dcolumn}
\usepackage{float}
\usepackage{bm}
\usepackage{amsmath}
\usepackage{mathrsfs}
\usepackage{slashed}
\usepackage{graphicx}
\begin{document}
\bibliographystyle{plain}

\newcommand*{\PKU}{School of Physics and State Key Laboratory of Nuclear Physics and
Technology, Peking University, Beijing 100871, China}\affiliation{\PKU}
\newcommand*{\chep}{Center for High Energy Physics, Peking University, Beijing 100871, China}\affiliation{\chep}

\title{ Superluminal Neutrinos in the Minimal Standard Model Extension}
\author{Nan Qin}\affiliation{\PKU}
\author{Bo-Qiang Ma}\email{mabq@pku.edu.cn}\affiliation{\PKU}\affiliation{\chep}

\begin{abstract}
The measurement of the neutrino velocity with the OPERA detector in
the CNGS beam shows unexpected indication, that the muon neutrino
velocity, $v_{\nu}$, exceeds the velocity of light in the vacuum,
$c$, which is obviously in contradiction with the most basic
hypothesis of modern physics. Within the framework of minimal
Standard Model Extension, we discuss the modified dispersion
relation and consequently the velocity-energy relation of muon
neutrinos. The simplified models are fitted to the OPERA data,
Fermilab experiment and MINOS data. We find that minimal Standard
Model Extension can describe these long baseline superluminal
neutrinos to a good accuracy. For the well-known tension between the
OPERA measurement and the Supernova 1987A neutrino observation, we
discuss two ways out of the contradiction.

\end{abstract}


\maketitle

\section{Superluminal Neutrinos in OPERA, Fermilab and MINOS Experiments}
Recently the OPERA neutrino experiment at the underground Gran Sasso
Laboratory reported their measurement of the velocity of neutrinos
from the CERN CNGS beam over a baseline of about
730~km~\cite{Adam2011}. Compared to the time taken for neutrinos
traveling at the speed of light in vacuum, an early arrival time of
$(60.7\pm6.9~(\mathrm{stat.})\pm 7.4~(\mathrm{sys.}))$~ns was
measured. This anomaly corresponds to a relative difference of the
muon neutrino velocity with respect to the speed of light
\begin{align}
\frac{v_{\nu}-c}{c}=(2.48\pm0.28~(\mathrm{stat.})\pm0.30~(\mathrm{sys.}))\times10^{-5},\label{opera}
\end{align}
at a significance of $6\sigma$. The OPERA result has already
inspired many papers with various discussions and
models~\cite{Ma:2011jj}. This is largely because there are
previously some supportive results from other collaborations. The
first direct measurement of neutrino velocity has been performed at
Fermilab long ago~\cite{Alspector1976,Kalbfleisch1979}. Based on
9,800 events, measurements of the velocity of muon neutrinos with
energy ranging from $30$~GeV to $200$~GeV gave
\begin{align}
|\beta_{\nu}-1|<10^{-5},\label{fermilab}
\end{align}
where $\beta_{\nu}\equiv v_{\nu}/c$. A few years ago, by using the
NuMI neutrino beam, the MINOS collaboration analyzed a total of 473
far detector neutrino events with an average energy $~3
$~GeV~\cite{Adamson2007}. They reported a shift with respect to the
expected time of flight of
\begin{align}
\delta_t=-126\pm32~(\mathrm{stat.})\pm64~(\mathrm{sys.)}~\mathrm{ns}
\end{align}
which corresponds to a constraint on the muon neutrino velocity,
\begin{align}
\frac{v_{\nu}-c}{c}=(5.1\pm2.9)\times10^{-5}
\end{align}
at $68\%$ confidence level. This $1.8\sigma$ signal was considered
consistent with zero, therefore it does not provide a strong
evidence of Lorentz violation effects. However, with the new
measurement of OPERA detector, it is surprising to notice that the
MINOS results and the OPERA results are compatible.

Although these measurements have been largely debated, the neutrino
velocity anomaly appears to be a strong challenge to the well-known
fact in special relativity that no physical signal travels faster
than the light. Special relativity, or Lorentz symmetry, is the
profound symmetry of space-time and has been incorporated into the
two cornerstones of modern physics: General Relativity and Quantum
Field Theory. However, the possible Lorentz symmetry violation (LV)
effects are sought for decades from various species of the standard
model, motivated by the unknown underlying theory of quantum gravity
together with various phenomenological
applications~\cite{Kostelecky2011,Shao2010,Shao2010a,Shao2011}. This
can happen in many alternative theories, e.g., the doubly special
relativity (DSR)~\cite{Amelino-Camelia2002,Zhang2011}, torsion in
general relativity~\cite{LV-GR1,LV-GR2}, and large
extra-dimensions~\cite{Ammosov2000,Pas2005}. In this paper, we will
work in an effective field theory framework, the Standard-Model
Extension (SME), in which Lorentz violation terms constructed with
Standard Model (SM) fields and controlling coefficients are added to
the usual SM lagrangian~\cite{Colladay1997,Colladay1998}. The
origins for such LV operators are suggested in many ways, of which
spontaneous Lorentz symmetry breaking proposed first in string
theory is widely recognized~\cite{Kostelecky1989,Kostelecky1991}.
There is also a recent proposal to derive some supplementary LV
terms from standard model with a basic consideration of the physical
invariance with respective to the mathematical background
manifolds~\cite{Ma10,SMS3}, and such a framework has been applied to
discuss the Lorentz violation effects for the cases of Dirac
particles~\cite{Ma10}, photons~\cite{SMS3,Ma10graal,SMS-photon2},
and neutrinos~\cite{SMS-OPERA}, in which the superluminal neutrinos
as a signal of Lorentz violation was suggested. In fact, the
possibilities of superluminal neutrinos were proposed with an
earlier version of SME~\cite{Chodos85} and with
extra-dimensions~\cite{Pas2005}. Here we do not get into such
discussions on theoretical details, but just analyze the experiments
of superluminal neutrinos with the effective lagrangian of SME given
in the next section.

\section{SME in Neutrino Sector and the Dispersion Relation}
The SME lagrangian in neutrino sector takes the
form~\cite{Colladay1998,Kostelecky2001,Yang2009}
\begin{align}
\mathcal{L}=\frac{1}{2}i\overline{\nu}_{A}\gamma^{\mu}\overleftrightarrow{D_{\mu}}\nu_{_B}\delta_{_{AB}}
+\frac{1}{2}ic_{_{AB}}^{\mu\nu}\overline{\nu}_{_A}\gamma_{\mu}\overleftrightarrow{D_{\nu}}\nu_{_B}
-a_{_{AB}}^{\mu}\overline{\nu}_{_A}\gamma_{\mu}\nu_{_B}+\cdots\;,\label{lagrangian}
\end{align}
where $c_{AB}^{\mu\nu}$ and $a_{AB}^{\mu}$ are Lorentz violation
coefficients resulting from tensor vacuum expectation values in the
underlying theory, the subscripts $A,B$ are flavor indices, and the
ellipsis denotes the non-renormalizable operators (eliminated in the
minimal SME). The first term in Eq.~(\ref{lagrangian}) is exactly
the SM operator, the second and third terms (CPT-even and CPT-odd
respectively) describe the contribution from Lorentz violation. For
simplicity, we neglect the effects due to interactions between
neutrinos and matters in which the neutrino beam propagates, and
thus replace $D_{\mu}$ with $\partial_{\mu}$ in
Eq.~(\ref{lagrangian}). Then after a simple transformation, we
arrive at
\begin{align}
\mathcal{L}=i\overline{\nu}_{A}\gamma^{\mu}\partial_{\mu}\nu_{_B}\delta_{_{AB}}
+ic_{_{AB}}^{\mu\nu}\overline{\nu}_{_A}\gamma_{\mu}\partial_{\nu}\nu_{_B}
-a_{_{AB}}^{\mu}\overline{\nu}_{_A}\gamma_{\mu}\nu_{_B}.\label{lag}
\end{align}
With Eq.~(\ref{lag}), we can easily get the Euler-Lagrange equation for neutrinos as
\begin{align}
(i\gamma^{\mu}\partial_{\mu}\delta_{AB}+c^{\mu\nu}_{AB}\gamma_{\mu}\partial_{\nu}-a^{\mu}_{AB}\gamma_{\mu})\nu_B=0.
\end{align}
One would find that the neutrino mass terms are missing here. This
is because such terms contribute to the neutrino velocity in the
form of $\left(\frac{m}{E}\right)^2$, totally negligible when GeV
neutrinos are discussed. Following the procedure presented in
Appendix.~A of Ref.~\cite{AlanKostelecky2004}, we arrive at the
effective hamiltonian as
\begin{align}
\newcommand{\DF}[2]{{\displaystyle\frac{#1}{#2}}}
 (H_{\mathrm{eff}})_{_{AB}}= \begin{pmatrix}
|\overrightarrow{p}|\delta_{_{AB}}+a_{_{AB}}^{\mu}\DF{p_{\mu}}{|\overrightarrow{p}|}-c_{_{AB}}^{\mu
\nu}\DF{p_{\mu}p_{\nu}}{|\overrightarrow{p}|}
& 0\\0 & |\overrightarrow{p}|\delta_{_{AB}}-a_{_{AB}}^{\mu}\DF{p_{\mu}}{|\overrightarrow{p}|}-c_{_{AB}}^{\mu
\nu}\DF{p_{\mu}p_{\nu}}{|\overrightarrow{p}|}
\end{pmatrix},\label{hamiltonian}
\end{align}
which is a $6\times6$ matrix when three generations of neutrinos are
considered, and the up-diagonal matrix denotes the hamiltonian for
neutrinos while the down-diagonal matrix denotes the hamiltonian for
anti-neutrinos. By diagonalizing this hamiltonian, one can get the
eigenenergies, the mixing matrix and consequently oscillation
probabilities of neutrinos. However, we are just interested in the
velocity or the dispersion relation of neutrinos, thus the
oscillation effects can be neglected in a first approximation.
Therefore, the model is simplified by including only one flavor,
i.e., $\nu_{\mu}$, so Eq.~(\ref{hamiltonian}) reduces to the
dispersion relation of the muon neutrinos as
\begin{align}
E=|\overrightarrow{p}|+\frac{1}{|\overrightarrow{p}|}(a^{\mu}p_{\mu}-c^{\mu\nu}p_{\mu}p_{\nu}).\label{dispersion}
\end{align}
With this dispersion relation, we could deduce the energy dependence
of the neutrino velocity. By comparing with experiment results
presented in the previous section, we can also fit the Lorentz
violation coefficients.

\section{Fits to Muon Neutrino Velocity Measurements}
We hold the same criterion when selecting data with that of
Ref.~\cite{Amelino-Camelia2011}. Data are collected from the
measurements of the muon neutrino velocity of
Fermilab~\cite{Alspector1976,Kalbfleisch1979},
MINOS~\cite{Adamson2007}, and OPERA~\cite{Adam2011} (even data of
muon anti-neutrinos from Fermilab are abandoned to keep things as
clean as possible). For Fermilab data, as noticed in their paper,
there might be a potential bias $b={b_0}^{+\sigma_b^+}_{-\sigma_b^-}
= 5^{+2}_{-1} \times 10^{-5}$ to the measured $v_{\nu}-c$. Hence
before using them, we corrected the bias, and the error bars are
squaredly summed, i.e., $\sigma_{\mathrm{new}}^2 =
\sigma_{\mathrm{old}}^2 + \sigma_b^2$. The bias-corrected data are
listed in Table~\ref{data}. For MINOS results, though the energy
spectrum for neutrinos has a long high-energy tail extending to
$120$~GeV, we pick the peak value $3$~GeV, which might induce some
unknown bias, but while the error bar is large for this data point,
its contribution to fitting is relatively small. For OPERA data, for
our purpose, we only use the $\nu_\mu$ CC interactions occurring in
the OPERA target, in total of 5489 events. When divided into two
bins with a separation at $20$~GeV, data produce results $v_{\nu}-c
= (2.17 \pm 0.83) \times 10^{-5}$ for low energy events with an
average energy $\left<E\right> = 13.9$~GeV, and $v_{\nu}-c = (2.74
\pm 0.80) \times 10^{-5}$ for high energy events with an average
energy $\left<E\right>=42.9$~GeV.

\begin{table}[H]
\caption{Data used for fitting from Fermilab~\protect\cite{Alspector1976,Kalbfleisch1979}, MINOS~\protect\cite{Adamson2007}, and OPERA~\protect\cite{Adam2011}.}
{\begin{tabular}{l|c|cccccccc}
\hline\hline Fermilab & ~~Energy (GeV)~~ & ~~32~~ & ~~44~~ & ~~59~~ &
~~69~~ & ~~90~~ & ~~120~~ & ~~170~~ & ~~195~~ \\ & $v_{\nu}-c$
($10^{-5}$) & $-2^{+2}_{-3}$ & $2\pm7$ & $-1^{+2}_{-3}$ &
$-1^{+2}_{-3}$ & $1^{+3}_{-4}$ & $1\pm7$ & $1^{+2}_{-3}$ &
$6^{+3}_{-4}$ \\ \hline MINOS & ~~Energy (GeV)~~ & ~~3~~ & & & & & & &
\\ & $v_{\nu}-c$ ($10^{-5}$) & $5.1\pm{2.9}$ & & & & & & &
\\ \hline OPERA & ~~Energy (GeV)~~ & ~~13.9~~ & ~~42.9~~ & & & & & &
\\ & $v_{\nu}-c$ ($10^{-5}$) & $2.17\pm0.83$ & $2.74\pm0.80$ & & &
& & & \\ \hline
\end{tabular} \label{data}}
\end{table}

Within our data preparation, there might be some unknown bias from
MINOS data and OPERA data, but the bias is assumed to be small.

Now we look back to our dispersion relation Eq.~(\ref{dispersion}),
in which there are totally $20$ unknown parameters since $\mu$ and
$\nu$ goes from $0$ to $3$. Fortunately the number is largely
reduced when certain symmetries are required in the model. For
instance, if we want the theory to satisfy rotational invariance,
non-diagonal entries in $c^{\mu\nu}$ and $a^i$ $(i=1,2,3)$ vanish
and there are only three nonzero parameters $a^0$, $c^{00}$ and
$c^{ii}=d$ $(i=1,2,3)$. Moreover, we can require the CPT invariance
to be an accurate symmetry thus get rid of the CPT-odd terms. Below
we analyze the data in two simple cases.

\noindent {\bf Case 1: All Lorentz violation coefficients but
$c^{00}$ vanish.}

\begin{figure}[H]
\begin{center}
\includegraphics[width=8cm]{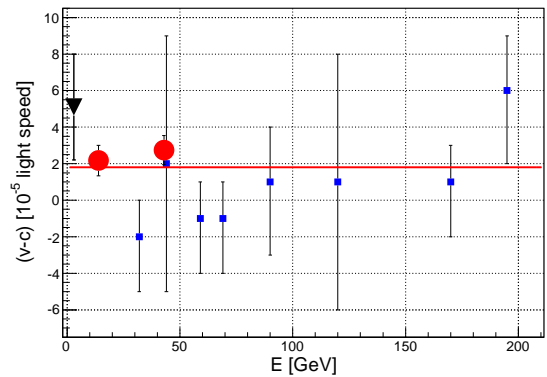}
\caption{The fit of a
  LV-modified velocity-energy relation, $v_{\nu} =\frac{\sqrt{4c^{00}+1}-1}{2c^{00}}$,
  to the moun neutrino velocity measurements from
  Fermilab~\protect\cite{Alspector1976,Kalbfleisch1979} (squares),
  MINOS~\protect\cite{Adamson2007} (triangles), and OPERA~\protect\cite{Adam2011}
  (circles).\label{c1}}
\end{center}
\end{figure}

In such a case the model is obviously invariant under CPT transformation and rotations. The dispersion relation reduces to
\begin{align}
E=|\overrightarrow{p}|-\frac{1}{|\overrightarrow{p}|}c^{00}E^2.\label{dis1}
\end{align}
By using the definition of the group velocity $v\equiv\frac{dE}{d|\overrightarrow{p}|}$ one can easily get the velocity for neutrinos as
\begin{align}
v_\nu=\frac{\sqrt{4c^{00}+1}-1}{2c^{00}},
\end{align}
which is a constant. This is easy to understand since all
$c^{\mu\nu}$s are dimensionless hence $E$ is proportional to
$|\overrightarrow{p}|$. So the energy dependence of the velocity
would not be changed if we include $c^{ii}=d\neq0$ in the dispersion
relation Eq.~(\ref{dis1}). The fit result is illustrated in
Fig.~(\ref{c1}) and the Lorentz violation coefficient $c^{00}$ is
constrained as $(-1.80\pm0.48)\times10^{-5}$.

\noindent {\bf Case 2:  $a^{0}\neq0$, $c^{00}\neq0$ while others
vanish.}

\begin{figure}[H]
\begin{center}
\includegraphics[width=8cm]{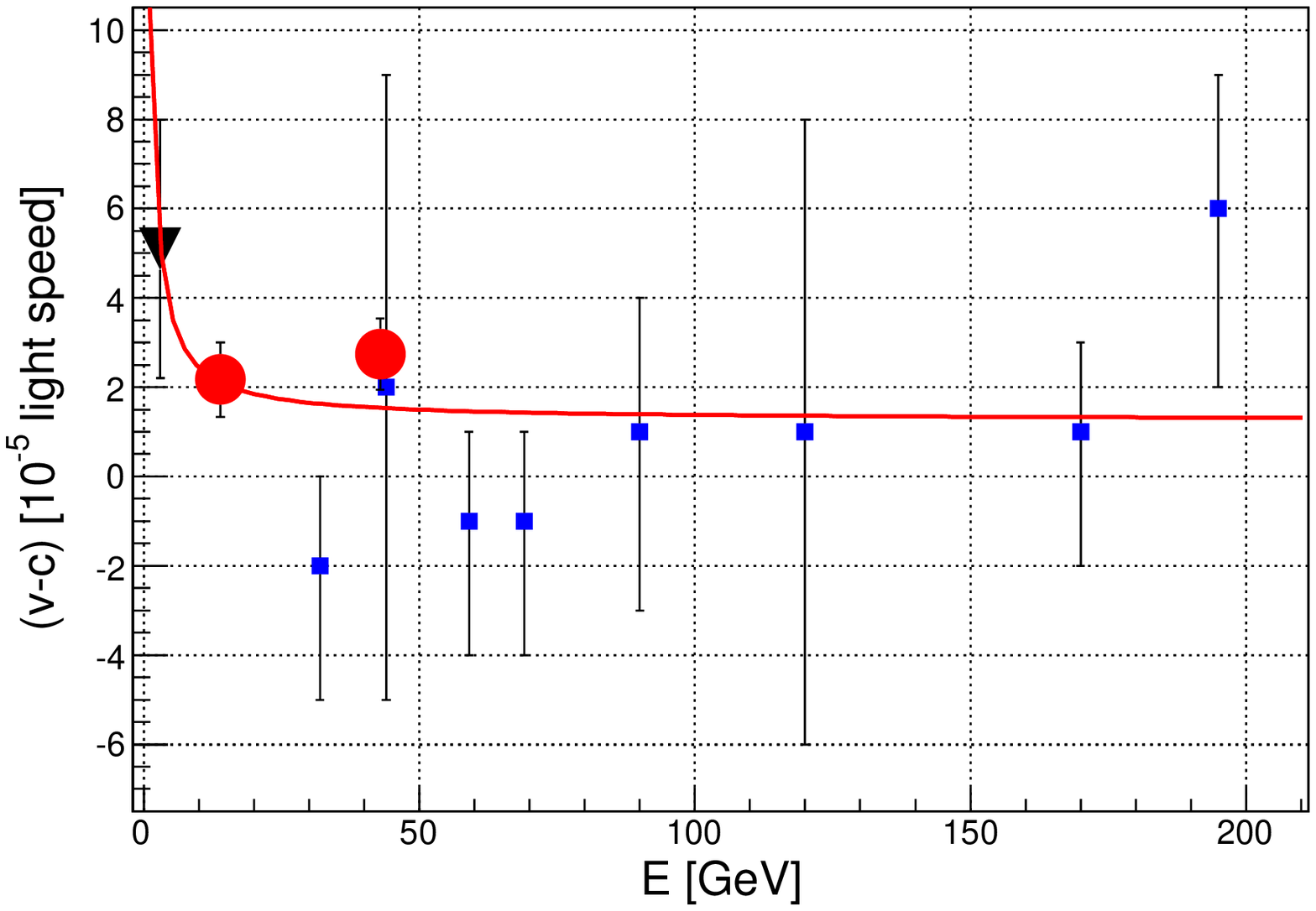}
\caption{The fit of a
  LV-modified velocity-energy relation,\protect\\ $v_{\nu} =\frac{\sqrt{E (-4 a^0+4 c^{00} E+E)}-2 a^0-(4 c^{00}+1) E}{\sqrt{E (-4 a^{0}+4 c^{00} E+E)}+2 a^0+4 c^{00} E+E}$,
  to the moun neutrino velocity measurements from
  Fermilab~\protect\cite{Alspector1976,Kalbfleisch1979} (squares),
  MINOS~\protect\cite{Adamson2007} (triangles), and OPERA~\protect\cite{Adam2011}
  (circles).\label{c2}}
\end{center}
\end{figure}

The CPT invariance is spoiled by a nonzero $a^0$ but the rotational symmetry still holds. The velocity-energy relation is given by
\begin{align}
v_\nu=\frac{\sqrt{E (-4 a^0+4 c^{00} E+E)}-2 a^0-(4 c^{00}+1) E}{\sqrt{E (-4 a^{0}+4 c^{00} E+E)}+2 a^0+4 c^{00} E+E}.
\end{align}
The fit result is illustrated in Fig.~(\ref{c2}) and the Lorentz
violation coefficients are constrained as
$c^{00}=(-1.26\pm0.63)\times10^{-5}$ and
$a^0=(-5.89\pm4.42)\times10^{-5}$~GeV~. Despite the large uncertainties
of $a^0$, we find in Fig.~(\ref{c2}) that the contribution from
CPT-odd terms becomes important in the low energy area. When energy
approaches to above $50$~GeV, the total shift of the neutrino speed
with respect to light speed due to CPT-even terms dominates.

There are many other choices of parameters and the fit results maybe
quite distinctive. From the dimensional analysis one can find
approximately that, CPT-even terms shift the velocity as a whole,
and CPT-odd terms modify the relation between energy and velocity.

\section{Supernova 1987A}
Besides the long baseline experiment, there are also measurable phenomenologies of superluminal neutrinos in astrophysics. For instance, supernova explosion (SNe) is an extremely
luminous event, which causes a burst of radiation that outshines an
entire galaxy. The radiation includes photons in a board range of
spectrum, as well as neutrinos. Actually, most energy of a SNe is
released in the form of neutrinos, however, due to the weak
interactions of neutrinos with matters, only one event is observed
with neutrino emissions on 23 February 1987, 7:35:35 UT ($\pm1$ min)
--- the Supernova 1987A in the Large Magellanic
Cloud~\cite{Hirata1987,Bionta1987}, which is optically observed on
24 February 1987. More than ten neutrinos were recorded with a
directional coincidence within the location of supernova explosion,
several hours before the optical lights are observed. Because of
weak interactions, neutrinos leak out of the dense environment
produced by the stellar collapse before the optical depth of photons
becomes $<1$. Hence an early-arrival of neutrinos is expected. The
journey of propagation of photons and neutrinos are of astrophysical
distance ($\sim 51.4$~kpc), hence it provides a unique opportunity
to measure the speed of neutrinos to be within the light speed with
a precision $\sim2\times10^{-9}$~\cite{Longo1987}.

Interestingly, while the OPERA results seem to be in remarkable
consistence with other terrestrial muon neutrino velocity
measurements, they contradict with the Supernova 1987A neutrino
observation severely. The two toy models in the previous section
obviously can not describe this contradiction. One way out is taking
more non-zero Lorentz coefficients into account, which will lead to
very complex velocity-energy dependence and spoil some global
symmetries. For instance, non-diagonal entries of $c^{\mu\nu}$ will
violate the rotational invariance. Moreover, we will need much more
data to test such models because of the large parameter space.

Another remedy is the observation that actually the species of
supernova neutrinos are different from those of terrestrial
neutrinos --- the former being electron neutrinos (and/or electron anti-neutrinos), while the measured neutrinos we are discussing 
are muon neutrinos. We suspect a family hierarchy should be 
responsible forthe observed different
velocities~\cite{SMS-OPERA,flavor-difference}. In the dispersion
relations, Lorentz violation coefficients of different flavors are
generally different, hence if there exist family hierarchies of
these parameters, the different propagation behaviors of Supernova
1987A neutrinos and terrestrial muon neutrinos can be understood.

It was argued by Cohen and Glashow~\cite{Glashow11} that the
energy-losing process $\nu_\mu\rightarrow\nu_\mu+e^{+}+e^{-}$
becomes kinematically allowed when the Lorentz violation of OPERA
muon neutrinos is of order $10^{-5}$, thus one would not see the muon
neutrinos at the LNGS, where the OPERA experiment was performed. Bi
{\it et al.} also argued that the Lorentz violation of muon
neutrinos of order $10^{-5}$ will forbid kinematically the
production process of muon neutrinos $\pi\rightarrow \mu + \nu_\mu$
for muon neutrinos with energy larger than about 5~GeV~\cite{Bi11}.
This kind of arguments has been recognized as a refutation for the
rationality of the OPERA experiment. However, there have been a
number of discussions~\cite{Smolin11,SMS-OPERA2,Li:2011ad},
indicating that such an argument is not valid in general. The
derived dispersion relation in the field theory frameworks could be
covariant with the momentum of the muon neutrino and thus can avoid
the Cherenkov-like radiations~\cite{Ma:2011jj,SMS-OPERA2}. The
framework of SME has the potential to accommodate both the
superluminality of neutrinos without the analogues Cherenkov
radiation. This work might be considered as a first estimate of the
magnitudes of the LV parameters in SME to fit the OPERA data, and
more investigations are still needed from both experimental and
theoretical aspects.

\section{Conclusion}
The OPERA group measured the difference between the velocity of muon
neutrinos to that of light, and reported a $6\sigma$ significant
indication that the muon neutrinos might travel with a speed
slightly larger than that of light, which obviously contradicts with
the most basic hypothesis underlying modern physics. Though unknown
systematical errors can exist potentially, it still seems extremely
worthy to look into possible theoretical reasons behind the
observations. Lorentz-violating-induced modified dispersion relation
appears to be the most robust possibility. With modified dispersion
relation, the velocity of muon neutrinos can depend on their
energies.

Within the framework of Minimal Standard Model Extension, we get the
modified dispersion relation. From the dispersion relation
Eq.~(\ref{dispersion}), energy dependence of neutrino velocity is
deduced in two rotational invariant models. We combine the muon
neutrino velocity measurements from Fermilab, MINOS, and OPERA, to
look into possible energy-dependence of neutrino velocity. The fit
results imply that a constant shift from the speed of light, with a
magnitude of order $10^{-5}c$ is the contribution of CPT-even terms,
while the energy dependence is due to CPT-odd terms.

We also explain the apparent conflicts between the Supernova 1987A
neutrino observation and the muon neutrino velocity measurements. We
point out that such a contradiction maybe solved by adding more
parameters at the cost of symmetry breaking, or the contradiction
maybe ascribe to family hierarchy of the Lorentz violation
coefficients. Either way, more data are needed to test the models.

\section*{Acknowledgements} We acknowledge the helpful discussions
with Lijing Shao, Zhi Xiao, Xinyu Zhang and Lingli Zhou. The work
was supported by National Natural Science Foundation of China
(Nos.~10975003, 11021092, 11035003 and 11120101004) and by the
Research Fund for the Doctoral Program of Higher Education (China).

\section*{Note Added}
There is a piece of news that the OPERA collaboration has identified two
possible effects that could have an influence on its neutrino timing
measurement. The first possible effect concerns an oscillator used
to provide the time stamps for GPS synchronizations, and the second
concerns the optical fibre connector that brings the external GPS
signal to the OPERA master clock. The two effects could have led to
an underestimate of the flight time of the neutrinos, and a
re-measurement of the neutrino speed by the OPERA collaboration will
be done in the near future. If this is the reason for the earlier
arrival time of neutrinos at the OPERA detector, the fitting result
of this paper will need some modification and updating. Within the
framework of SME, neutrinos can either be superluminal or subluminal
depending on the sign of the parameters, so we still need more data
to make a conclusion.


\end{document}